\documentclass[aps,pra,showpacs,superscriptaddress,preprint]{revtex4}
\usepackage{graphicx}

\catcode`ð=\active
 \defð{\u{g}}
 \catcode`Ð=\active
 \defÐ{\u{G}}
 \catcode`Ý=\active
 \defÝ{\. I}
 \catcode`ö=\active
 \defö{\"{o}}
 \catcode`Ö=\active
 \defÖ{\"O}
 \catcode`ü=\active
 \defü{\"{u}}
 \catcode`Ü=\active
 \defÜ{\"{U}}
 \catcode`Þ=\active
 \defÞ{\c{S}}
 \catcode`þ=\active
 \defþ{\c{s}}
 \catcode`ý=\active
 \defý{{\i}}
 \catcode`ç=\active
 \defç{\d{c}}
 \catcode`Ç=\active
 \defÇ{\d{C}}

\begin{document}

\title{Approximate Analytical Solutions of the Pseudospin Symmetric Dirac Equation for
Exponential-type Potentials}

\author{\small Altuð Arda}
\email[E-mail: ]{arda@hacettepe.edu.tr}\affiliation{Department of
Physics Education, Hacettepe University, 06800, Ankara,Turkey}
\author{\small Ramazan Sever}
\email[E-mail: ]{sever@metu.edu.tr}\affiliation{Department of
Physics, Middle East Technical  University, 06800, Ankara,Turkey}
\author{\small Cevdet Tezcan}
\email[E-mail: ]{ctezcan@baskent.edu.tr}\affiliation{Faculty of
Engineering, Baþkent University, Baglýca Campus, Ankara,Turkey}

\date{\today}

\begin{abstract}
The solvability of The Dirac equation is studied for the
exponential-type potentials with the pseudospin symmetry by using
the parametric generalization of the  Nikiforov-Uvarov method. The
energy eigenvalue equation, and the corresponding Dirac spinors
for Morse, Hulthen, and $q$-deformed Rosen-Morse potentials are
obtained within the framework of an approximation to the
spin-orbit coupling term, so the solutions are
given for any value of the spin-orbit quantum number $\kappa=0$, or $\kappa \neq 0$.
\end{abstract}

\maketitle

\newpage

\section{Introduction}
The concept of pseudo-spin was constructed firstly in spherical
nuclei [1, 2], and it is observed experimentally that the single
particle levels labeled as pseudo-spin doublets are very close in
energy [3]. The pseudo-spin doublets in nuclei are decomposed by
using radial ($n_r$), orbital ($\ell$\,), and total angular
momentum ($j$) quantum numbers as ($n_r, \ell, j=\ell+1/2$), and
($n_{r}-1, \ell+2, j=\ell+3/2$). The pseudo orbital angular
momentum, $\widetilde{\ell}=\ell+1$, and the pseudo spin,
$\widetilde{s}$, quantum numbers give total angular momentum,
$j=\widetilde{\ell}+\widetilde{s}$, and the pseudo-spin doublets
are degenerate with respect to pseudo spin, $\widetilde{s}$ [4].
The pseudo-spin doublets occur in nuclei, when the magnitude of
scalar, $V_s(r)$, and vector, $V_v(r)$, potentials are nearly
equal, with opposite sign, i.e., $V_s(r) \simeq-V_v(r)$ [5]. The
pseudo-spin symmetry is studied based on Dirac equation in real
nuclei, and shown that it is related with the competition between
the centrifugal barrier, and pseudo-spin orbital potential [6].
The pseudo-spin concept is discussed in deformed nuclei [7], and
exotic nuclei as well [3]. It is observed that the pseudo-spin
symmetry is also an important one in the case of triaxiality [8].

The pseudo-spin orbital potential creates the splitting of the
pseudo-spin doublets, and the pseudo-spin symmetry is exact
symmetry in real nuclei, when the derivative of the difference
between scalar, and vector potentials vanishes, but the above
condition gives a good symmetry for exotic nuclei [3, 6]. The
pseudo-spin symmetry is identified as a $SU(2)$ symmetry of the
Dirac Hamiltonian, under the condition that the sum of scalar, and
vector potentials is equal to zero [9]. Recently, it is pointed
out that the shape of the lower components of the Dirac spinor for
the doublets is the same, when the pseudo-spin doublets are
degenerate [4]. It is showed that it becomes possible to construct
a map which relates the normal state ($\ell, s$) with the pseudo
state ($\widetilde{\ell}, \widetilde{s}$) by applying of the
helicity operator to the non-relativistic single-particle
eigenfunction to understand of the mechanism to generate the
pseudo-spin symmetry [10]. The pseudo-spin concept is discussed in
the non-relativistic harmonic oscillator, and obtained that the
condition between the coefficients of spin-orbit, and orbit-orbit
terms in the case of non-relativistic single-particle Hamiltonian
having the pseudo-spin symmetry is consistent with the result
obtained relativistic mean-field theory [11, 12].

The pseudo-spin symmetry concept has found a great application
area, especially in nuclear theory. The identical bands observed
in nuclei are explained by using the pseudo-spin symmetry [13].
The idea of the pseudo-spin has been considered to be useful to
construct an effective shell-model coupling scheme [14]. Some
features of nuclei, such as deformation, and superdeformation can
be explained in the context of the pseudo-spin symmetry [13, 15,
16].

Recently, the solutions of the Klein-Gordon, and Dirac equations
including the spin-orbit coupling term have been studied by many
authors for different potentials, such as Morse potential [17-20],
P\"{o}schl-Teller potential [21-23], Woods-Saxon potential [24],
Eckart [25-28], harmonic oscillator [29, 30], three parameter
diatomic molecular potential [31], and angle-dependent potential
[32]. In the present work, we deal with the approximate solutions,
and corresponding wave functions of the Dirac equation including
spin-orbit coupling term under the exact pseudo-spin symmetry for
Morse, Hulth\'{e}n, and $q$-deformed Rosen-Morse potentials. We
point out that the parametric generalization of the
Nikiforov-Uvarov method can be applied to the Dirac equation with
the above potentials, and the energy eigenvalue equation, and
corresponding eigenfunctions can be obtained for the values of
spin-orbit quantum number $\kappa=0$, or $\kappa\neq0$.

\section{The Dirac Equation with Spin-Orbit Coupling}
The Dirac equation for a fermion with mass $m$ moving in an
external scalar, and vector potentials reads ($\hbar=c=1$)

\begin{eqnarray}
[\alpha\,.\,\hat{p}+\beta[m+V_s(r)]+V_v(r)]\Psi(r)=E\Psi(r)\,,
\end{eqnarray}
where $E$ is the energy of the particle, $\hat{p}$ is the
three-momentum operator, and $\alpha$, and $\beta$ are the
$4\times4$ Dirac matrices written in terms of$2\times2$ Pauli
matrices, and unit matrix. Under the consideration that the system
has a spherical symmetry for which the potential fields depend on
the radial coordinate, the quantum state  of the particle is
labeled by the quantum number set ($n_{r}, j, m, \kappa$), where
$m$ is the projection of the total angular momentum on the
$z$-axis, and $\kappa=\pm(j+1/2)$ is the eigenvalues of the
operator $\hat{\kappa}=-\beta(\hat{\sigma}\,.\,\hat{L}+1$) [33].
Here, $\kappa=-(j+1/2)$ denotes the aligned spin ($s_{1/2},
p_{3/2}, etc.$), and $\kappa=+(j+1/2)$ denotes the unaligned spin
($p_{1/2}, d_{3/2}, etc.$). The spherically symmetric Dirac wave
function can than be written in terms of upper, and lower
components as [34]

\begin{eqnarray}
\Psi(r)=\,\frac{1}{r}\,\Bigg(\begin{array}{c}
  \,f\,(r)[\,Y_{\ell}\,\chi\,]_{m}^{j} \\
ig\,(r)[\,Y_{\tilde{\ell}}\,\chi\,]_{m}^{j}
\end{array}\Bigg)\,
\end{eqnarray}
where $f(r)$, and $g(r)$ are the radial wave functions,
$Y_{\ell}\,(\theta,\phi)$, and $\chi$ are the spherical, and spin
functions, respectively. Substituting Eq (2) into Eq. (1), we get
the following radial Dirac equations

\begin{eqnarray}
\Bigg(\,\frac{d}{dr}\,+\,\frac{\kappa}{r}\,\Bigg)\,f(r)-[M(r)+\epsilon\,]g(r)=0\,,\\
\Bigg(\,\frac{d}{dr}\,-\,\frac{\kappa}{r}\,\Bigg)\,g(r)-[M(r)-\epsilon\,]f(r)=0\,.
\end{eqnarray}
where $M(r)=m+V_s(r)$, and $\epsilon=E-V_v(r)$. Using the
expression for $g(r)$ obtained from Eq. (3), and inserting it into
Eq. (4), we have two second order differential equations including
spin-orbit coupling term

\begin{eqnarray}
\Big\{\,\frac{d^2}{dr^2}-\,\frac{\kappa(\kappa+1)}{r^2}\,-[M^2(r)-\epsilon^2]
\Big\}f(r)=\Big\{\frac{1}{M(r)+\epsilon}\frac{d}{dr}[V_s(r)-V_v(r)](\frac{d}{dr}\,+\,\frac{\kappa}{r})
\Big\}f(r)\,,\\
\Big\{\,\frac{d^2}{dr^2}-\,\frac{\kappa(\kappa-1)}{r^2}\,-[M^2(r)-\epsilon^2]
\Big\}g(r)=\Big\{\frac{1}{M(r)-\epsilon}\frac{d}{dr}[V_s(r)+V_v(r)](\frac{d}{dr}\,-\,\frac{\kappa}{r})
\Big\}g(r)\,,
\end{eqnarray}

Under the condition of the exact pseudo-spin symmetry, i.e.,
$\frac{d}{dr}[V_v(r)+V_s(r)]=0$, or $V_v(r)+V_s(r)=C=const.$, Eq.
(6) gives

\begin{eqnarray}
\Big\{\,\frac{d^2}{dr^2}-\,\frac{\kappa(\kappa-1)}{r^2}\,+[m-E+C][V_v(r)-V_s(r)]\Big\}\,g(r)
=[m^2-E^2+C(m+E)]g(r)\,.
\end{eqnarray}
From the last equation, the energy eigenvalues depend also on the
quantum number $\tilde{\ell}$ because of the relations given by
$\kappa(\kappa-1)=\tilde{\ell}(\tilde{\ell}+1)$, and
$\kappa(\kappa+1)=\ell(\ell+1)$. So, the energy eigenstates with
$j=\tilde{\ell}\pm1/2$ are degenerate for $\tilde{\ell}\neq0$,
which gives the situation of the exact pseudo-spin symmetry in the
Dirac equation.

In the present work, we intend to solve the last equation four
different potentials, Morse, Hulth\'{e}n, and $q$-deformed
Rosen-Morse potentials, namely. The Dirac equation in Eq. (7) can
not be solved exactly because of the spin-orbit coupling term. So,
we use the Pekeris approximation [35] to find the suitable
expression instead of the spin-orbit coupling term in the case of
Morse potential. This approximation makes possible to write the
interaction term in the form of the Morse potential in terms of
new parameters $D_i(i=0, 1, 2)$. In the case of Hulth\'{e}n, and
$q$-deformed Rosen-Morse potential, we introduce the following
approximation instead of the spin-orbit coupling term [22]

\begin{eqnarray}
\frac{1}{r^2} \approx \frac{\alpha^2 e^{-\alpha r}}{(1-e^{-\alpha
r})^2}\,,
\end{eqnarray}
which gives a second order differential equation from Eq. (7)
without $(1/r^2)$-term. In all case of potentials, we use the
parametric generalization of the Nikiforov-Uvarov method [36], and
so we prove that the parametric version of the method can be
applied to the Dirac equation with Morse, P\"{o}schl-Teller,
Hulth\'{e}n, and $q$-deformed Rosen-Morse potential.

\section{Nikiforov-Uvarov Method}

By using an appropriate coordinate transformation, the
Schr\"{o}dinger equation is transformed into the following form

\begin{eqnarray}
\Psi''(s)+\frac{\tilde{\tau}(s)}{\sigma(s)}
\Psi'(s)+\frac{\tilde{\sigma}(s)}{{\sigma}^{2}(s)}\Psi(s)=0\,,
\end{eqnarray}
where $\sigma(s)$, $\tilde{\sigma}(s)$ are polynomials, at most
second degree, and $\tilde{\tau}(s)$ is a first degree polynomial.
In the NU-method, the polynomial $\pi(s)$, and the parameter $k$
are required, and defined as

\begin{eqnarray}
\pi(s)=\frac{1}{2}\,[\sigma^{\prime}(s)-\tilde{\tau}(s)]\pm
\sqrt{\frac{1}{4}\,[\sigma^{\prime}(s)-\tilde{\tau}(s)]^2-
\tilde{\sigma}(s)+k\sigma(s)},
\end{eqnarray}
and

\begin{eqnarray}
\lambda=k+\pi^{\prime}(s ),
\end{eqnarray}
where $\lambda$ is a constant. The function under the square root
in the polynomial in $\pi(s)$ in Eq. (10) must be square of a
polynomial in order that $\pi(s)$ be a first degree polynomial, so
the derivative of $\pi(s)$ is a constant, and this defines the
constant $k$. Replacing $k$ into Eq. (10), we define

\begin{eqnarray}
\tau(s)=\tilde{\tau}(s)+2\pi(s).
\end{eqnarray}
where the derivative  of $\tau(s)$ should be negative [35], which
let us know the choice of the solution. The hypergeometric type
equation in Eq. (9) has a particular solution with degree $n$, if
$\lambda$ in Eq. (11) satisfies

\begin{eqnarray}
\lambda=\lambda_{n}=-n\tau^{\prime}-\frac{\left[n(n-1)\sigma^{\prime\prime}\right]}{2},
\quad n=0,1,2,\ldots
\end{eqnarray}
To obtain the solution of Eq. (9) it is assumed that the solution
is a product of two independent parts

\begin{eqnarray}
\Psi(s)=\phi(s)~y(s),
\end{eqnarray}
where $y(s)$ can be written as

\begin{eqnarray}
y_{n}(s)= \frac{a_{n}}{\rho(s)}\frac{d^{n}}{ds^{n}}
\left[\sigma^{n}(s)~\rho(s)\right],
\end{eqnarray}
where $a_{n}$ is normalization constant, and the function
$\rho(s)$ is the weight function, and should satisfy the condition

\begin{eqnarray}
\frac{d\sigma(s)}{ds}\rho(s)+\sigma(s)\frac{d\rho(s)}{ds}=\tau(s)~\rho(s)\,,
\end{eqnarray}
The other factor is defined as

\begin{eqnarray}
\frac{\phi^{\prime}(s)}{\phi(s)}=\frac{\pi(s)}{\sigma(s)}.
\end{eqnarray}
In order to clarify the parametric generalization of the NU
method, let us take the following equation, which represents a
general form of the Schr\"{o}dinger-like equation written for any
potential,

\begin{eqnarray}
\left[\frac{d^{2}}{ds^{2}}+\frac{\alpha_{1}-\alpha_{2}s}{s(1-\alpha_{3}s)}
\frac{d}{ds}+\frac{-\xi_{1}s^{2}+\xi_{2}s-\xi_{3}}{[s(1-\alpha_{3}s)]^{2}}\right]\Psi(s)=0.
\end{eqnarray}
When Eq. (18) is compared with Eq. (9), we get

\begin{eqnarray}
\tilde{\tau}(s)=\alpha_{1}-\alpha_{2}s\,\,\,;\,\,\sigma(s)=s(1-\alpha_{3}s)\,\,\,;\,\,
\tilde{\sigma}(s)=-\xi_{1}s^{2}+\xi_{2}s-\xi_{3}\,.
\end{eqnarray}
Substituting these into Eq. (10)

\begin{eqnarray}
\pi(s)=\alpha_{4}+\alpha_{5}s\pm\sqrt{(\alpha_{6}-k\alpha_{3})s^{2}+(\alpha_{7}+k)s+\alpha_{8}}\,,
\end{eqnarray}
where the parameter set are

\begin{eqnarray}
\begin{array}{ll}
\alpha_{4}=\frac{1}{2}\,(1-\alpha_{1})\,, & \alpha_{5}=\frac{1}{2}\,(\alpha_{2}-2\alpha_{3})\,, \\
\alpha_{6}=\alpha_{5}^{2}+\xi_{1}\,, &
\alpha_{7}=2\alpha_{4}\alpha_{5}-\xi_{2}\,, \\
\alpha_{8}=\alpha_{4}^{2}+\xi_{3}\,. & \\
\end{array}
\end{eqnarray}
In NU-method, the function under the square root in Eq. (20) must
be the square of a polynomial [35]. This condition gives the roots
of the parameter $k$, and they can be written as

\begin{eqnarray}
k_{1,2}=-(\alpha_{7}+2\alpha_{3}\alpha_{8})\pm2\sqrt{\alpha_{8}\alpha_{9}}\,,
\end{eqnarray}
where the $k$-values can be real or imaginary, and
$\alpha_{9}=\alpha_{3}\alpha_{7}+\alpha_{3}^{2}\alpha_{8}+\alpha_{6}$\,.
Different $k$'s lead to the different $\pi(s)$'s. For

\begin{eqnarray}
k=-(\alpha_{7}+2\alpha_{3}\alpha_{8})-2\sqrt{\alpha_{8}\alpha_{9}}\,,
\end{eqnarray}
$\pi(s)$ becomes

\begin{eqnarray}
\pi(s)=\alpha_{4}+\alpha_{5}s-\left[(\sqrt{\alpha_{9}}+\alpha_{3}\sqrt{\alpha_{8}}\,)s-\sqrt{\alpha_{8}}\,\right]\,,
\end{eqnarray}
and also

\begin{eqnarray}
\tau(s)=\alpha_{1}+2\alpha_{4}-(\alpha_{2}-2\alpha_{5})s-2\left[(\sqrt{\alpha_{9}}
+\alpha_{3}\sqrt{\alpha_{8}}\,)s-\sqrt{\alpha_{8}}\,\right].
\end{eqnarray}
Thus, we impose the following for satisfying the condition that
the derivative of the function $\tau(s)$ should be negative in the
method

\begin{eqnarray}
\tau^{\prime}(s)&=&-(\alpha_{2}-2\alpha_{5})-2(\sqrt{\alpha_{9}}+\alpha_{3}\sqrt{\alpha_{8}}\,)
\nonumber \\
&=&-2\alpha_{3}-2(\sqrt{\alpha_{9}}+\alpha_{3}\sqrt{\alpha_{8}}\,)\quad<0.
\end{eqnarray}
From Eqs. (11), (12), (25), and (26), and equating Eq. (11) with
the condition that $\lambda$ should satisfy given by Eq. (13), we
find the eigenvalue equation

\begin{eqnarray}
\alpha_{2}n-(2n+1)\alpha_{5}&+&(2n+1)(\sqrt{\alpha_{9}}+\alpha_{3}\sqrt{\alpha_{8}}\,)+n(n-1)\alpha_{3}\nonumber\\
     &+&\alpha_{7}+2\alpha_{3}\alpha_{8}+2\sqrt{\alpha_{8}\alpha_{9}}=0.
\end{eqnarray}
We get from Eq. (16)

\begin{eqnarray}
\rho(s)=s^{\alpha_{10}-1}(1-\alpha_{3}s)^{\frac{\alpha_{11}}{\alpha_{3}}-\alpha_{10}-1}\,,
\end{eqnarray}
and substituting into Eq. (15) gives

\begin{eqnarray}
y_{n}(s)=P_{n}^{(\alpha_{10}-1,\frac{\alpha_{11}}{\alpha_{3}}-\alpha_{10}-1)}(1-2\alpha_{3}s)\,,
\end{eqnarray}
where

\begin{eqnarray}
\alpha_{10}=\alpha_{1}+2\alpha_{4}+2\sqrt{\alpha_{8}}\,\,;\,
\alpha_{11}=\alpha_{2}-2\alpha_{5}+2(\sqrt{\alpha_{9}}+\alpha_{3}\sqrt{\alpha_{8}})\,.
\end{eqnarray}
and $P_{n}^{(\alpha,\beta)}(1-2\alpha_{3}s)$ are the Jacobi
polynomials. From Eq. (17), one gets

\begin{eqnarray}
\phi(s)=s^{\alpha_{12}}(1-\alpha_{3}s)^{-\alpha_{12}-\frac{\alpha_{13}}{\alpha_{3}}}\,,
\end{eqnarray}
then the general solution $\Psi(s)=\phi(s)y(s)$ becomes

\begin{eqnarray}
\Psi(s)=s^{\alpha_{12}}(1-\alpha_{3}s)^{-\alpha_{12}-\frac{\alpha_{13}}{\alpha_{3}}}
P_{n}^{(\alpha_{10}-1,\frac{\alpha_{11}}{\alpha_{3}}-\alpha_{10}-1)}(1-2\alpha_{3}s)\,
\end{eqnarray}
where

\begin{eqnarray}
\alpha_{12}=\alpha_{4}+\sqrt{\alpha_{8}}\,\,;\,\alpha_{13}=\alpha_{5}-(\sqrt{\alpha_{9}}+\alpha_{3}\sqrt{\alpha_{8}}\,)\,.
\end{eqnarray}

\section{Bound-State Solutions}
\subsection{Morse Potential}
We set the difference between scalar, and vector potentials in Eq.
(6) as the Morse potential [37] given by

\begin{eqnarray}
V_{Morse}(r)=D[e^{-2\alpha x}-2e^{-\alpha x}]\,,
\end{eqnarray}
where $D$ is the dissociation energy of the potential,
$\alpha=ar_0$, $x=\frac{r}{r_0}-1$, $r_0$ is the equilibrium
distance, and the positive parameter $a$ is the potential width.
According to the Pekeris approximation, the spin-orbit coupling
term can be written in terms of three new parameters $D_0, D_1$,
and $D_2$ as

\begin{eqnarray}
\frac{\kappa(\kappa-1)}{r^2}\simeq\gamma(D_0+D_1 e^{-\alpha x}+D_2
e^{-2\alpha x})\,,
\end{eqnarray}
where $\gamma=\kappa(\kappa-1))r^2_0$ (see Ref [17], for details).
Substituting Eq. (34), and (35) into Eq. (6), we get

\begin{eqnarray}
\frac{d^2g(x)}{dx^2}+\Big\{a^2_3 e^{-2\alpha x}-a^2_2 e^{-\alpha
x}-a^2_1\Big\}g(x)=0\,,
\end{eqnarray}
where $a^2_1=r^2_0[\gamma D_0+\Sigma_{m}(m+E)]$,
$a^2_2=r^2_0(\gamma D_1+2D\Sigma_{m})$, and
$a^2_3=r^2_0(D\Sigma_{m}-\gamma D_2)$ in the above equations. By
using the new variable $s=e^{-\alpha x}$, we obtain

\begin{eqnarray}
\frac{d^2g(s)}{ds^2}+\frac{1}{s}\frac{dg(s)}{ds}\,+\,\Big\{a^2_3\beta^2-\frac{a^2_2\beta^2}{s}-
\frac{a^2_1\beta^2}{s^2}\Big\}g(s)=0\,.
\end{eqnarray}
where $\beta^2=1/\alpha^2$, and $\Sigma_m=m-E+C$. Comparing the
Eq. (37) with Eq. (18), we obtain the following parameter set

\begin{eqnarray}
\begin{array}{ll}
\alpha_1=1\,, & -\xi_1=a^2_3\beta^2 \\
\alpha_2=0\,, &
\xi_2=-a^2_2\beta^2 \\
\alpha_3=0\,, &
-\xi_3=-a^2_1\beta^2 \\ \alpha_4=0\,, & \alpha_5=0 \\
\alpha_6=\xi_1\,, & \alpha_7=-\xi_2 \\
\alpha_8=\xi_3\,, & \alpha_9=\xi_1 \\
\alpha_{10}=1+2\sqrt{\xi_3\,}\,, & \alpha_{11}=2\sqrt{\xi_1\,} \\
\alpha_{12}=\sqrt{\xi_3}\,, & \alpha_{13}=-\sqrt{\xi_1}
\end{array}
\end{eqnarray}
Substituting the above expressions in Eq. (27), we obtain the
energy eigenvalue equation of a fermion moving in the Morse
potential under the exact pseudo-spin symmetry

\begin{eqnarray}
(2n+1)\sqrt{\gamma D_2-D\Sigma_{m}\,}&+&\beta r_0(\gamma
D_1+2D\Sigma_m)\nonumber\\&+&2\beta r_0\sqrt{[\gamma
D_0+\Sigma_{m}(m+E)](\gamma D_2-D\Sigma_m)\,}=0\,,
\end{eqnarray}
and the lower spinor component from Eq. (32)

\begin{eqnarray}
g(s)&=&s^{\beta r_0\sqrt{\gamma D_0+\Sigma_{m}(m+E)\,}}e^{-\beta
r_0\sqrt{\gamma D_2-D\Sigma_m\,}\,s}\nonumber\\&\times&L^{2\beta
r_0\sqrt{\gamma D_0+\Sigma_{m}(m+E)\,}}_n\,(2\beta r_0\sqrt{\gamma
D_2-D\Sigma_m\,}s)\,,
\end{eqnarray}
where $L^{(k)}_n(x)$ are the Laguerre polynomials.

We give the upper spinor component from Eq. (4), by using Eq. (40)

\begin{eqnarray}
f(s)&=&a\Sigma^{-1}_m s^{\beta r_0\sqrt{\gamma
D_0+\Sigma_{m}(m+E)\,}}e^{-\beta r_0\sqrt{\gamma
D_2-D\Sigma_m\,}\,s}\nonumber\\&\times&\Bigg\{\beta
r_0\sqrt{\gamma D_2-D\Sigma_m\,}s\Bigg[L^{2\beta r_0\sqrt{\gamma
D_0+\Sigma_{m}(m+E)\,}}_n\,(2\beta r_0\sqrt{\gamma
D_2-D\Sigma_m\,}s)\nonumber\\&+&L^{1+2\beta r_0\sqrt{\gamma
D_0+\Sigma_{m}(m+E)\,}}_{n-1}\,(2\beta r_0\sqrt{\gamma
D_2-D\Sigma_m\,}s)\Bigg]\nonumber\\&-&\Bigg[\beta r_0\sqrt{\gamma
D_0+\Sigma_{m}(m+E)\,}+\frac{\kappa}{ar_0 -ln
s}\Bigg]\nonumber\\&\times&L^{2\beta r_0\sqrt{\gamma
D_0+\Sigma_{m}(m+E)\,}}_n\,(2\beta r_0\sqrt{\gamma
D_2-D\Sigma_m\,}s)\Bigg\}\,.
\end{eqnarray}
where we use some recursion relations of the Laguerre polynomials.

\subsection{Hulth\'{e}n Potential}
The Hulth\'{e}n potential reads

\begin{eqnarray}
V(r)=-V_0\frac{e^{-x}}{1-e^{-x}}\,,
\end{eqnarray}
where $x=\frac{r}{r_0}$, and $r_0$ denotes the spatial range [38,
39]. We choose the difference between scalar, and vector
potentials as Hulth\'{e}n potential potential, and using the
approximation in Eq. (8), which becomes $\frac{1}{r^2}\simeq
\frac{e^{x}}{r_0^2(e^{x}-1)^2}$ in the present case, we obtain
from Eq. (6)

\begin{eqnarray}
\frac{d^2g(r)}{dr^2}-\Bigg\{\frac{\kappa(\kappa-1)e^{x}}{r_0^2(e^{x}-1)^2}+\Sigma_m\Big(
m+E+\frac{V_0}{e^{x}-1}\Big)\Bigg\}g(r)=0\,,
\end{eqnarray}
Using the new variable $s=e^{-x}$, Eq. (43) becomes

\begin{eqnarray}
\frac{d^2g(s)}{ds^2}&+&\frac{1-s}{s(1-s)}\frac{dg(s)}{ds}+\frac{1}{[s(1-s)]^2}
\Bigg\{-r^2_0\Sigma_{m}(m+E)\nonumber\\&+&\Big[-\kappa(\kappa-1)+2r^2_0\Sigma_{m}(m+E)
-r^2_0\Sigma_{m}V_0\Big]s\nonumber\\&+&\Big[-r^2_0\Sigma_{m}(m+E)+r^2_0\Sigma_{m}V_0\Big]s^2\Bigg\}g(s)=0\,.
\end{eqnarray}
Comparing with Eq. (18), we get

\begin{eqnarray}
\begin{array}{ll}
\alpha_1=1\,, & -\xi_1=r^2_0\Sigma_{m}[-m-E+V_0] \\
\alpha_2=1\,, &
\xi_2=r^2_0\Sigma_{m}[2(m+E)-V_0]-\kappa(\kappa-1) \\
\alpha_3=1\,, &
-\xi_3=-r^2_0\Sigma_{m}(m+E) \\ \alpha_4=0\,, & \alpha_5=-1/2 \\
\alpha_6=\frac{1}{4}+\xi_1\,, & \alpha_7=-\xi_2 \\
\alpha_8=\xi_3\,, & \alpha_9=\xi_1-\xi_2+\xi_3+\frac{1}{4} \\
\alpha_{10}=1+2\sqrt{\xi_3\,}\,, & \alpha_{11}=2+2(\sqrt{\xi_1-\xi_2+\xi_3+\frac{1}{4}\,}+\sqrt{\xi_3\,}\,) \\
\alpha_{12}=\sqrt{\xi_3}\,, &
\alpha_{13}=-\frac{1}{2}-\sqrt{\xi_1-\xi_2+\xi_3+\frac{1}{4}\,}-\sqrt{\xi_3}
\end{array}
\end{eqnarray}

The energy eigenvalue equation is written from Eq. (27)

\begin{eqnarray}
r_0\sqrt{\Sigma_{m}(m+E)\,}\Big[2n+1+\sqrt{4\kappa(\kappa-1)+1\,}\,\Big]+(n+\frac{1}{2}\,)
\sqrt{4\kappa(\kappa-1)+1\,}\nonumber\\+n(n+1)+\kappa(\kappa-1)+r^2_0\Sigma_{m}V_0+\frac{1}{2}=0\,,
\end{eqnarray}
and corresponding lower, and upper Dirac spinors by using Eqs. (4)
and (32), respectively,

\begin{eqnarray}
g(s)&=&s^{r_0\sqrt{\Sigma_{m}(m+E)}}\,(1-s)^{\frac{1}{2}[1+\sqrt{4\kappa(\kappa-1)+1\,}\,]}
\nonumber\\&\times&P^{(2r_0\sqrt{\Sigma_{m}(m+E)\,}\,,\,\sqrt{4\kappa(\kappa-1)+1\,}\,)}_n(1-2s)\,,\\
f(s)&=&(r_0\Sigma_{m})^{-1}\,s^{r_0\sqrt{\Sigma_{m}(m+E)}}(1-s)^{\frac{1}{2}[1+\sqrt{4\kappa(\kappa-1)+1\,}\,]}
\nonumber\\&\times&\Big\{\Big[-r_0\sqrt{\Sigma_{m}(m+E)}+\frac{1}{2}[1+\sqrt{4\kappa(\kappa-1)+1\,}\,]\frac{s}{1-s}
\nonumber\\&+&\frac{\kappa}{lns}\Big]P^{(2r_0\sqrt{\Sigma_{m}(m+E)\,}\,,\,\sqrt{4\kappa(\kappa-1)+1\,}\,)}_n(1-2s)
\nonumber\\&+&\Big(n+2r_0\sqrt{\Sigma_{m}(m+E)}+\sqrt{4\kappa(\kappa-1)+1\,}+1\Big)s\,\nonumber\\&\times&
P^{(1+2r_0\sqrt{\Sigma_{m}(m+E)\,}\,,\,1+\sqrt{\kappa(\kappa-1)+1\,}\,)}_{n-1}(1-2s)\,.
\end{eqnarray}
where $P^{(k,m)}_n(x)$ are the Jacobi polynomials, and we have
used some recursion relations of the Jacobi polynomials to obtain
Eq. (48).

\subsection{$q$-Deformed Rosen-Morse Potential}
The $q$-deformed Rosen-Morse Potential reads [40]

\begin{eqnarray}
V(r)=\frac{V_1}{1+qe^{-2x}}-V_2 q\frac{e^{-2x}}{(1+qe^{-2x})^2}\,,
\end{eqnarray}
where $x=\alpha r$, and $q$ is the deformation parameter.

We prefer to solve the Dirac equation for $\kappa=0$ in Eq. (6)
with the $PT$-symmetric version of the potential by setting to the
difference between scalar, and vector potentials

\begin{eqnarray}
V^{PT}(r)=\frac{V_1}{1+qe^{-2ix}}-V_2
q\frac{e^{-2ix}}{(1+qe^{-2ix})^2}\,.
\end{eqnarray}
We obtain the following equation under the above consideration by
using the transformation $s=-e^{-2ix}$

\begin{eqnarray}
\frac{d^2g(s)}{ds^2}&+&\frac{1-qs}{s(1-qs)}\frac{dg(s)}{ds}+\frac{1}{[s(1-qs)]^2}\Big\{
\delta^2[V_1
\Sigma_{m}-\Sigma_{m}(m+E)]\nonumber\\&+&\delta^2\Sigma_{m}[2q(m+E)-qV_1+V_2]s-\delta^2q^2\Sigma_{m}
(m+E)s^2\Big\}g(s)=0\,,
\end{eqnarray}
Comparing the Eq. (51) with Eq. (18), we get

\begin{eqnarray}
\begin{array}{ll}
\alpha_1=1\,, & -\xi_1=\delta^2q^2\Sigma_{m}
(m+E) \\
\alpha_2=q\,, &
\xi_2=\delta^2\Sigma_{m}[2q(m+E)-qV_1+V_2] \\
\alpha_3=q\,, & -\xi_3=\delta^2[V_1
\Sigma_{m}-\Sigma_{m}(m+E)] \\ \alpha_4=0\,, & \alpha_5=-q/2 \\
\alpha_6=\frac{q^2}{4}+\xi_1\,, & \alpha_7=-\xi_2 \\
\alpha_8=\xi_3\,, & \alpha_9=\xi_1-q\xi_2+q^2\xi_3+\frac{q^2}{4} \\
\alpha_{10}=1+2\sqrt{\xi_3\,}\,, & \alpha_{11}=2q+2(\sqrt{\xi_1-q\xi_2+q^2\xi_3+\frac{^2}{4}\,}+q\sqrt{\xi_3\,}\,) \\
\alpha_{12}=\sqrt{\xi_3}\,, &
\alpha_{13}=-\frac{q}{2}-\sqrt{\xi_1-q\xi_2+q^1\xi_3+\frac{q^2}{4}\,}+q\sqrt{\xi_3}
\end{array}
\end{eqnarray}

The energy eigenvalue equation is written from Eq. (27)

\begin{eqnarray}
2\delta\sqrt{\Sigma_{m}(m+E-V_1)\,}\Big[2n+1+\frac{1}{2}\,\sqrt{1-(4V_2\Sigma_{m})/q\,}\,\Big]
+(n+\frac{1}{2})\sqrt{1-(4V_2\Sigma_{m})/q\,}\nonumber\\+n(n+1)+\frac{1}{2}-\delta^2\Sigma_{m}(V_1+V_2)/q=0\,.
\end{eqnarray}
and corresponding lower, and upper Dirac spinors by using Eqs. (4)
and (32), respectively,

\begin{eqnarray}
g(s)&=&s^{\delta\sqrt{\Sigma_{m}(m+E-V_1)}}\,(1-qs)^{\frac{1}{2}[1+\sqrt{1-(4V_2\Sigma_{m})/q\,}\,]}
\nonumber\\&\times&P^{(2\delta\sqrt{\Sigma_{m}(m+E-V_1)\,}\,,\,\sqrt{1-(4V_2\Sigma_{m})/q\,}\,)}_n(1-2qs)\,,\\
f(s)&=&2i\alpha
\Sigma_{m}^{-1}\,s^{\delta\sqrt{\Sigma_{m}(m+E-V_1)}}(1-qs)^{\frac{1}{2}[1+\sqrt{1-(4V_2\Sigma_{m})/q\,}\,]}
\nonumber\\&\times&\Big\{\Big[-\delta\sqrt{\Sigma_{m}(m+E-V_1)}+\frac{q}{2}[1+\sqrt{1-(4V_2\Sigma_{m})/q\,}\,]\frac{s}{1-qs}
\nonumber\\&-&\frac{\kappa}{lns}\Big]P^{(2\delta\sqrt{\Sigma_{m}(m+E-V1)\,}\,,\,\sqrt{1-(4V_2\Sigma_{m})/q\,}\,)}_n(1-2qs)
\nonumber\\&+&q\Big(n+2\delta\sqrt{\Sigma_{m}(m+E-V_1)}+\sqrt{1-(4V_2\Sigma_{m})/q\,}+1\Big)\,\nonumber\\&\times&
P^{(1+2\delta\sqrt{\Sigma_{m}(m+E-V_1)\,}\,,\,1+\sqrt{1-(4V_2\Sigma_{m})/q\,}\,)}_{n-1}(1-2qs)\,.
\end{eqnarray}
where $\delta^2=1/(4\alpha^2)$ in the above equations.

\section{Conclusion}
We have approximately solved the Dirac equation for the Morse,
Hulth\'{e}n, and $q$-deformed Rosen-Morse potentials with the
exact pseudospin symmetry for arbitrary spin-orbit quantum number
$\kappa$. We have found the eigenvalue equation, and corresponding
Dirac spinors in terms of Jacobi (or Laguerre) polynomials by
using the parametric generalization of the NU-method within the
framework of an approximation to the spin-orbit coupling term. We
have observed that the parametric form of the NU method can be
used to solve the Dirac equation with the above potentials. Our
results for the cases of Morse potential is good agreement with
the ones obtained in the literature.

\section{Acknowledgments}
This research was partially supported by the Scientific and
Technical Research Council of Turkey.


\newpage

\begin{thebibliography}{99}

\bibitem{1} A.~Arima, M.~Harvey, and K.~Shimizu, Phys. Lett. B {\bf 30}, 517 (1969).

\bibitem{2} K.~T.~Hecht, and A.~Adler, Nucl. Phys. A {\bf 137}, 139 (1969).

\bibitem{3} J.~Meng, K.~Sugaware-Tanabe, S.~Yamaji, and A.~Arima, Phys.\ Rev.\ C {\bf 59}, 154
(1999).

\bibitem{4} J.~N.~Ginocchio, and D.~G.~Madland, Phys.\ Rev.\ C {\bf 57},
1167 (1998).

\bibitem{5} J.~N.~Ginocchio, Phys. Rev. Lett. {\bf 78}(3), 436 (1997).

\bibitem{6} J.~Meng, K.~Sugaware-Tanabe, S.~Yamaji, P.~Ring, and A.~Arima, Phys.\ Rev.\ C {\bf 58},
R628 (1998).

\bibitem{7} R.~D.~Ratna Raju, J.~P.~Draayer, and K.~T.~Hecht, Nucl. Phys. A {\bf 202}, 433 (1973);
J.~P.~Draayer, and K.~J.~Weeks, Ann. Phys. (N.Y.) {\bf 156}, 41
(1984).

\bibitem{8}  T.~Beuschel, A.~L.~Blokhin, and J.~P.~Draayer, Nucl. Phys. A {\bf 619}, 119 (1997);
A.~L.~Blokhin, T.~Beuschel, J.~P.~Draayer, and C.~Bahri, Nucl.
Phys. A {\bf 612}, 163 (1997).

\bibitem{9} J.~N.~Ginocchio, and A.~Leviatan, Phys. Lett. B {\bf 245}, 1 (1998).

\bibitem{10} A.~L.~Blokhin, C.~Bahri, and J.~P.~Draayer, Phys. Rev. Lett. {\bf 74}, 4149 (1995).

\bibitem{11} C.~Bahri, J.~P.~Draayer, and S.~A.~Moszkowski, Phys. Rev. Lett. {\bf 68}, 2133 (1992).

\bibitem{12} J.~N.~Ginocchio, Phys. Rep. {\bf 414}, 165 (2005).

\bibitem{13} A.~Bohr, I.~Hamamoto, and B.~R.~Mottelson, Phys. Scr. {\bf 26}, 267 (1982);
B.~Mottelson, Nucl. Phys. A {\bf 522}, 1 (1991).

\bibitem{14} D.~Troltenier, C.~Bahri, and J.~P.~Draayer, Nucl. Phys. A {\bf 586}, 53 (1995).

\bibitem{15} J.~Dudek, W.~Nazarewicz, Z.~Szymanski, and G.~Le Ander, Phys. Rev. Lett. {\bf 59}, 1405 (1987).

\bibitem{16} F.~S.~Stephens et \textit{al.}, Phys. Rev. Lett. {\bf 64}, 2623
(1990).

\bibitem{17} C.~Berkdemir, Nucl. Phys. A {\bf 770}, 32 (2006).

\bibitem{18}  W.~C.~Qiang, R.~S.~Zhou, and Y.~Gao, J. Phys. A: Math. Theor.
{\bf 40}, 1677 (2007).

\bibitem{19} O.~Bayrak, and I.~Boztosun, J. Phys. A: Math. Theor.
{\bf 40}, 11119 (2007).

\bibitem{20}  A.~Alhaidari, Phys. Rev. Lett. {\bf 87}, 210405
(2001).

\bibitem{21} C.~S.~Jia, P.~Guo, Y.~F.~Diao, L.~Z.~Yi, and X.~J.~Xie, Eur. Phys. J. A {\bf 34}, 41 (2007).

\bibitem{22} Y.~Xu, S.~He, and C.~S.~Jia, J. Phys. A {\bf 41}, 255302 (2008).

\bibitem{23} A.~Alhaidari, J. Phys. A {\bf 34}, 9287 (2001).

\bibitem{24} J.~Y.~Gou, and Z.~Q.~Sheng, Phys. Lett. A {\bf 338}, 90 (2005).

\bibitem{25} C.~S.~Jia, P.~Guo, and X.~L.~Pery, J. Phys. A: Math. Gen. {\bf 39}, 7737 (2006).

\bibitem{26} L.~H.~Zhang, X.~P.~Li, and C.~S.~Jia, Phys. Lett. A {\bf 372}, 2201 (2008).

\bibitem{27} A.~Soylu, O.~Bayrak, and I.~Boztosun, J. Phys. A: Math. Theor.
{\bf 41}, 065308 (2008).

\bibitem{28} A.~Alhaidari, J. Phys. A {\bf 37}, 5805 (2004).

\bibitem{29} R.~Lisboa, M.~Malheiro, A.~S.~De Castro, P.~Alberto, and M.~Fiolhais, Phys. Rev. C {\bf 69}, 024319 (2004).

\bibitem{30} A.~S.~De Castro, P.~Alberto, R.~Lisboa, and M.~Malheiro, Phys. Rev. C {\bf 73}, 054309 (2006).

\bibitem{31} C.~S.~Jia, J.~Y.~Liu, L.~He, and L.~T.~Sun, Phys. Scr. {\bf 75}, 388 (2007).

\bibitem{32} C.~Berkdemir, and R.~Sever, J. Phys. A: Math. Gen.
{\bf 41}, 045302 (2008).

\bibitem{33} W.~Greiner, B.~M\"{u}ller, and J.~Rafelski, Quantum Electrodynamics of Strong
Fields (Springer-Verlag, New Yok, 1985).

\bibitem{34} A.~Leviatan, and J.~N.~Ginocchio, Phys. Lett. B {\bf 518}, 214
(2001).

\bibitem{35} C.~L.~Pekeris, Phys. Rev. {\bf 45}, 98 (1934).

\bibitem{36} A.~F.~Nikiforov, and V.~B.~Uvarov, Special Functions of
Mathematical Physics (Birkhauser, Basel, 1988).

\bibitem{37} P.~M.~Morse, Phys. Rev. {\bf 34}, 57 (1929).

\bibitem{38} L.~Hulth\'{e}n, Ark. Mat. Astron. Fys. A {\bf 28}, 5 (1942).

\bibitem{39} C.-Y.~Chen, D.-S.~Sun, and F.-L.~Lu, Phys. Lett. A
{\bf 370}, 219 (2007).

\bibitem{40} C.~Tezcan, and R.~Sever, J. Math. Chem. {\bf 42}, 387, (2006);
C.~S.~Jia, Y.~Sun, and Y.~Li, Phys.\ Lett.\  A {\bf 331}, 115 (2003).
\end{thebibliography}
\end{document}